# Centroid Detection by Gaussian Pattern Matching in Adaptive Optics

Akondi Vyas
Indian Institute of Astrophysics
Koramangala, Bangalore, India
Indian Institute of Science
Koramangala, Bangalore, India
91-9986734022

vyas@iiap.res.in

M B Roopashree
Indian Institute of Astrophysics
Koramangala, Bangalore, India
91-9886070233

roopashree@iiap.res.in

B Raghavendra Prasad
Indian Institute of Astrophysics
Koramangala, Bangalore, India
91-9449597577

brp@iiap.res.in

## ABSTRACT
Shack Hartmann wavefront sensor is a two dimensional array of lenslets which is used to detect the incoming phase distorted wavefront through local tilt measurements made by recording the spot pattern near the focal plane. Wavefront reconstruction is performed in two stages - (a) image centroiding to calculate local slopes, (b) formation of the wavefront shape from local slope measurement. Centroiding accuracy contributes to most of the wavefront reconstruction error in Shack Hartmann sensor based adaptive optics system with readout and background noise. It becomes even more difficult in atmospheric adaptive optics case, where scintillation effects may also occur. In this paper we used a denoising technique based on thresholded Zernike reconstructor to minimize the effects due to readout and background noise. At low signal to noise ratio, this denoising technique can be improved further by taking the advantage of the shape of the spot. Assuming a Gaussian pattern for individual spots, it is shown that the centroiding accuracy can be improved in the presence of strong scintillations and background.

## Categories and Subject Descriptors
I.4.9 [**Image Processing and Computer Vision**]: Applications

## General Terms
Algorithms, Measurement, Performance, Experimentation

## Keywords
Adaptive Optics, Shack Hartmann Sensor, Wavefront Reconstruction, Centroiding

## 1. INTRODUCTION
Adaptive Optics (AO) is a well developed technology that helps in quality imaging of objects hidden behind turbulent media. It is widely used in astronomical imaging, retinal imaging and free space communication systems [1-3]. A simplest AO system comprises of a wavefront sensor that detects the shape of the incoming optical wavefront and a wavefront corrector that compensates the effects of wavefront distortions by imposing a conjugate wavefront on the incident wavefront [4]. The wavefront sensor and corrector are connected via a processor where the required wavefront reconstruction computations are performed. The information of the shape of the conjugated wavefront is then communicated to the corrector.

Although different wavefront measuring devices exist like curvature sensor, shearing interferometer, common path interferometer, most generally used sensor is the Shack Hartmann Sensor (SHS) which is made up of tiny lenses arranged in a two dimensional array. A wavefront incident on the SHS forms an array of spots near the focal plane of the lenses. Unlike an ideal plane wavefront normal to the SHS which forms well focused equidistant spots, a distorted wavefront forms spots that are not equidistant as shown in Fig. 1. From the measurement of the shift in the spot positions from ideal case, it is possible to calculate the local slopes (over each lenslet/subaperture) of the wavefront. Using the slope values, the approximate shape of the wavefront is estimated. The two steps involved in wavefront reconstruction are hence (a) centroiding of individual spots to calculate local slopes, (b) formation of the wavefront shape from local slope measurements.

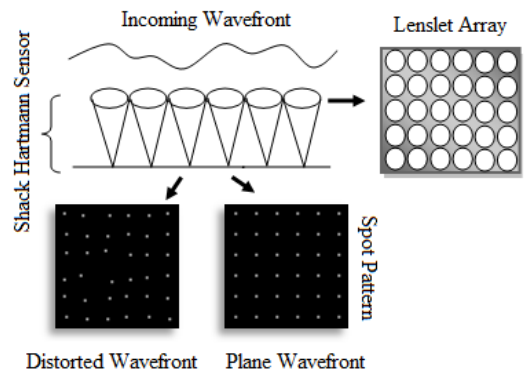

**Figure 1 Principle of Shack Hartmann Sensor**

Centroid detection becomes a difficult problem in the presence of noise [5]. Readout noise, background noise, photon noise and scintillation effects are dominant effects in atmospheric AO. Due to the presence of noise, mathematically simplest centroiding algorithm like Center of Gravity (CoG) method leads to large wavefront reconstruction errors [6]. To improve upon this technique, Weighted Center of Gravity (WCoG), Iteratively Weighted Center of Gravity (IWCoG), Intensity Weighted Centroiding (IWC), correlation based centroiding, matched filtering based centroiding were developed [7-9]. In this paper, we studied the performance of CoG, WCoG, IWCoG and IWC techniques. The centroid, $(x_c, y_c)$ can be computed in the case of CoG using the formula,

$$\left(x_c, y_c\right) = \frac{\sum_{ij} X_{ij} I_{ij}}{\sum_{ij} I_{ij}} \qquad (1)$$

where $I_{ij}$ is a two dimensional image function with M×M pixels, $i$ and $j$ take values 1,2,….M. In the case of WCoG, a weighting function (generally Gaussian) is used to take the advantage of the shape of the spot.

$$\left(x_c, y_c\right) = \frac{\sum_{ij} X_{ij} I_{ij} W_{ij}}{\sum_{ij} I_{ij} W_{ij}} \qquad (2)$$

$$W(x, y) = \frac{1}{2\pi\sigma^2} \exp\left\{-\frac{(x-x_c)^2}{2\sigma^2} - \frac{(y-y_c)^2}{2\sigma^2}\right\} \qquad (3)$$

where, σ is the spread of the spot.

IWCoG works on the same principle as WCoG, but with a difference that it is an iterative method where the center of the weighting function is shifted to the position of the estimated centroid in the previous iteration. In IWC, the weighting function is defined as powers of intensity, $W(x, y) = \{I(x, y)\}^p$, p is real and positive. CoG performs best in high Signal to Noise Ratio (SNR) conditions. WCoG is generally used in the closed loop AO where the shift in the spot is small. This technique shows a significant improvement over CoG in the presence of readout and background noise. IWCoG can be used effectively in closed as well as open loop AO systems but this technique is time consuming and has problems in iterative convergence. IWC performs best in photon noise limited conditions [10].

The phase values are related to the measured slopes through a linear system of equations, Ax=b, where 'A' is the coefficient matrix, 'x' is a vector containing phase values to be calculated and 'b' represents the measured slope values [11]. The coefficient matrix is a sparse matrix in AO case and hence sparse matrix methods are generally used for wavefront reconstruction from slope values [12].

There are many factors that cause wavefront reconstruction errors in AO like sensor alignment, sensor discretization, detector pixelization, detector SNR, detector readout noise, frame jitter, cross talk between lenslets, background light and scintillations effects due to large intensity fluctuations (also magnified while using a Laser Guide Star (LGS) as reference star), random spot wobbling during the exposure time and many other random effects [13]. It was shown theoretically and experimentally that the scintillations can greatly degrade the correction ability of AO systems [14]. Scintillations depend mainly on the Fried parameter which indicates the strength of turbulence and the variance of intensity [15].

A simple denoising technique based on Zernike reconstruction that greatly improves the wavefront reconstruction accuracy was used in this paper [16]. In this noise removal method the spot patterns corresponding to individual subapertures of a SHS are reconstructed using Zernike polynomials by calculating Zernike moments. The reconstructed images are thresholded and the resultant images are used for centroid estimation. Since the noise features arising due to background, readout are of low spatial extent, they will not be found on the images reconstructed using less number of Zernike moments. The scintillation effects or rather prominent events comparable to the actual spot image will sustain even after following the Zernike reconstruction and thresholding process.

In this paper we propose a Gaussian Pattern Matching (GPM) algorithm to improve the centroid detection accuracy. In this method we take the advantage of the shape of the spot pattern. As a first step, we identify the features. Secondly, we selectively eliminate the features that do not have a circular shape. Finally we recount the number of features, $f_N$ and if it is greater than one, we use the fact that the intensity distribution from the feature centroid drops like a Gaussian. This algorithm can be further improved by including wavefront prediction algorithms [17].

The methodology used for simulating the SHS spot pattern in the presence of readout noise, photon noise and background noise is described in section 2. Zernike reconstructor based denoising technique is presented in section 3 with illustrations and graphs. Section 4 describes the basic idea behind the Gaussian pattern matching algorithm and the design steps involved to minimize faulty recognition. The results obtained by applying Monte Carlo simulations on various centroiding algorithms are presented in section 5 and the derived conclusions are presented briefly in section 6.

## 2. SIMULATION OF SPOT PATTERN

Photon noise, background noise and detector readout noise are three major noise sources in AO. In astronomical adaptive optics system, a Shack Hartmann spot pattern is greatly influenced by photon noise which is caused due to the low photon count. Photon noise can be minimized by using a LGS or a bright Natural Guide Star (NGS) as a reference star. Background noise occurs because of the finite sky brightness and other unwanted light sources. Readout noise is an unavoidable noise caused by the detector. We included these effects in our simulations. The following steps were followed to simulate the spot pattern.

- A two dimensional Gaussian function was simulated on an array of size M×M pixels with control over the position at which the spot center lies and the spread of the spot. Sample ideal spot images are shown in Fig. 2.

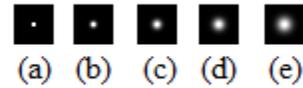

**Figure 2 Gaussian spots with a spread, σ taking values (a) 0.5 (b) 1.0 (c) 1.5 (d) 2.0 (e) 2.5 (image center is the spot center)**

- A photon noise limited image is then simulated as a second step. Photon noise is an intensity dependent noise. In this step, each of the pixel values is replaced by a randomly chosen value from a Poisson distribution whose mean is the actual pixel value. The actual spot and the photon noise limited spot images are shown in Fig. 3. The pixel values were amplified for better visualization.
- Background noise which is caused due to unwanted light sources is generally random. This can be simulated by generating pseudo Gaussian random numbers. Choosing a suitable SNR, the maximum background intensity is adjusted. The effect of background noise on photon noise limited spot image with SNR=1.33 is shown in Fig. 4.

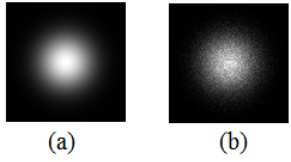

**Figure 3 Poisson Noise limited image (b) in comparison with an ideal Gaussian spot (a)**

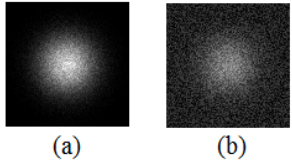

**Figure 4 (a) Photon noise limited (b) Background noise superimposed on photon noise limited image, SNR=1.33**

- Detector readout noise is projected as random fluctuations about the actual intensity value. This noise becomes critical during short exposure time scale imaging, at low light level conditions and faint background noise levels. This is due to the detector imperfection in signal detection. The impact of readout noise in low light level condition on the ideal Shack Hartmann spot pattern is similar to the effect of a uniform background noise.

The spot pattern hence obtained after addition of the above described noise is used for further analysis. They are used in the Monte Carlo simulations for testing the centroiding algorithms. In the simulations the user has control over the actual position of the spot. The Centroid Estimation Error (CEE) can be evaluated using the formula,

$$CEE = \sqrt{(x_c^* - x_c)^2 + (y_c^* - y_c)^2} \quad (4)$$

where, $(x_c^*, y_c^*)$ is the estimated centroid and $(x_c, y_c)$ is the actual centroid. Effectively, CEE measures the distance of the estimated centroid from the actual position of the spot.

## 3. ZERNIKE RECONSTRUCTOR + THRESHOLDING

The denoising technique of spot images consists of two steps. Firstly the images are represented in terms of a finite number of Zernike polynomials. The process of conversion of the image in terms of Zernike polynomials is performed by the Zernike reconstructor. Thresholding is performed on the reconstructed images to obtain a nearly noise free image.

### 3.1 Zernike Reconstructor

Zernike polynomials are a set of continuous orthogonal circular polynomials defined over the unit disk. Since they form a complete set of orthogonal polynomials, any two dimensional function, I($x$, $y$) can be represented as a proper linear combination of this basis set.

$$I(x, y) = \sum_i a_i Z_i(x, y) \quad (5)$$

where, $Z_i(x, y)$ represents Zernike polynomials and the coefficients $a_i$'s standing before Zernike polynomials are image weights called Zernike moments. Since Zernike polynomials are mathematically complicated functions, the calculation of Zernike moments of images is tedious. In this paper, we used a very fast and nearly accurate method suggested by Hosny to compute the Zernike moments [18].

The denoising ability varies with the number of Zernike orders used for representing the image. The reconstructed noisy spot pattern images are shown in Fig. 5. It is advised to use more number of Zernike moments since the central Gaussian feature can be captured better during reconstruction.

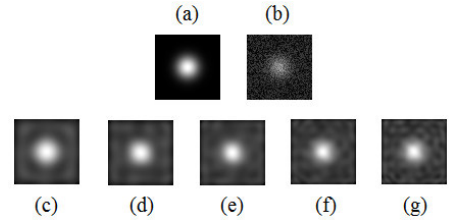

**Figure 5 Representation of images using Zernike moments (a) Ideal spot image (b) Noisy image. Zernike reconstructed image using (c) 10 (d) 15 (e) 20 (f) 25 (g) 30 moments.**

### 3.2 Thresholding

The image reconstructed using the above method is then thresholded to remove the higher order aberrations (low spatial extent events) on the spot pattern. It can be observed by a comparison of Fig. 5c and Fig. 5g. that lower spatial scale events will be reconstructed only if further more orders of Zernike polynomials are used. The images in Fig. 5c-g are thresholded and shown in Fig. 6.

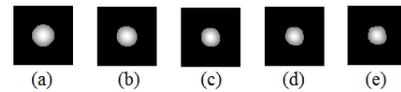

**Figure 6 Images formed after Zernike reconstruction and thresholding Figure 5b. using (a) 10 (b) 15 (c) 20 (d) 25 (e) 30 Zernike moments**

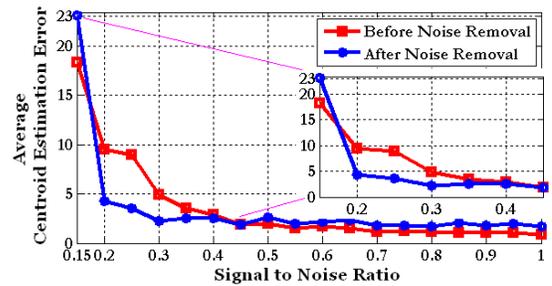

**Figure 7 Performance of IWCoG method in the presense of noise before and after noise removal algorithm**

This method is advantageous in the presence of very high noise level conditions as shown in Fig. 7. At SNR<0.45, the denoising procedure in conjugation with CoG method performs better than IWCoG with 8 iterations [16]. This helps is overcoming the convergence and speed problems associated with IWCoG.

### 3.3 Erroneous spot pattern after thresholding

At high noise conditions, the thresholded Zernike reconstructed images lead to multiple features as shown in Fig. 8. This is due to the fact that at high noise level conditions there can be large scale features (scales comparable to the size of the spot) which might not be removed even after the denoising procedure is implemented.

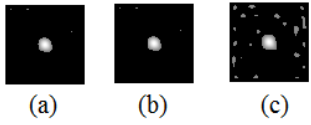

**Figure 8 Erroneous threholding at SNR=1 for percentage thresholding of (a) 80 (b) 70 (c) 60**

## 4. GAUSSIAN PATTERN MATCHING

To use this erroneous spot for accurate centroiding, we take the advantage of the fact that the final spot formed at the focal plane of a lens must assume an Airy pattern which can be approximated to a Gaussian like structure. The pattern matching algorithm is implemented in three steps:

- Feature recognition
- Shape identification
- Profile identification

### 4.1 Feature recognition

In this step, the features on the spot pattern image are counted and identified. Feature recognition can be performed by using many existing pattern recognition algorithms. In this paper, we used a simple Hough peak identification method to detect the features and number them in the order of peak height. To eliminate small scale features which may arise due to unavoidable scintillations, we imposed threshold conditions on the size of the features. The number of features varied from 30 to 1 from very low SNR to high SNR.

### 4.2 Shape identification

Most features do not have a circular shape as shown in Fig. 8c. The circularity or the extent of the feature being circular is measured for each of the features. In the proposed algorithm, the centroid position was computed for a single feature locally and the distance from the centriod at which intensity becomes zero is measured. This distance is called the radius parameter. The radius parameter is estimated at different angles (0-360) from the feature centroid position. We define circularity, C of a single feature as the inverse of the variance of the radius parameter computed over different angles from the centroid position. For an ideal circular feature, the variance is zero and hence the circularity is infinity. A lower cutoff for this parameter is chosen to eliminate features that are not close to a circular shape. This shape identification process was applied to Fig 8(c) and is shown in Fig. 9. It is possible to eliminate all the noise features in this step at better SNR and applying a suitable feature size thresholding.

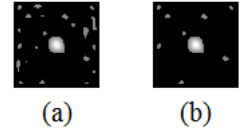

**Figure 9 Shape Identification: (a) Erroneous spot pattern image (b) Spot pattern after removal of small scale features and elimination of non-circular features**

### 4.3 Profile identification

In the previous step, the shape of the spot was used for selective elimination. There can be cases of large noise features that are circular in shape as shown in Fig. 10. In the profile identification step, the intensity profile is used to choose the actual spot feature. The number of features are recounted and re-identified after going through the shape identification step. The intensity fall off from the centroid position of individual features is measured and pattern matching was performed with a standard one dimensional Gaussian function, $G(x) = \exp[-(x-x_0)^2/2\sigma^2]$. In this process, the features that do not follow a Gaussian like structure are eliminated. A comparison of the actual spot line profile with the noise feature line profiles is shown in Fig. 11. The sum of squares of errors was used as a statistic to measure the total deviation of the actual values from the fitted values.

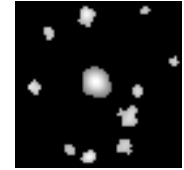

**Figure 10 Comparably large circular features**

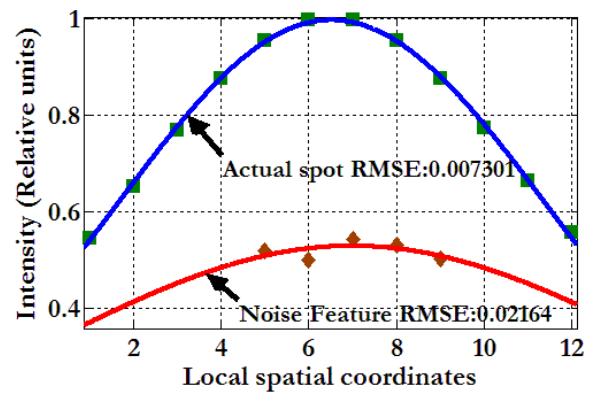

**Figure 11 Comparison of Gaussian fitted line profile for actual spot and a noisy feature**

## 5. COMPUTATIONAL RESULTS

Monte Carlo simulations were performed using the spots simulated as described in section 2. Zernike Reconstructor + Thresholding algorithm is implemented as described in section 3.

The images formed after thresholding are given as input to the Gaussian pattern matching (GPM) algorithm, where a single feature is selected from a set of features erroneously left on the spot pattern after thresholding. Simulations were run on 100 randomly simulated noisy spot pattern images (64×64) with the spot occupying 28×28 pixels. The performance of CoG and IWC before and after applying GPM algorithm on images processed using thresholded Zernike reconstruction is shown in Fig. 12.

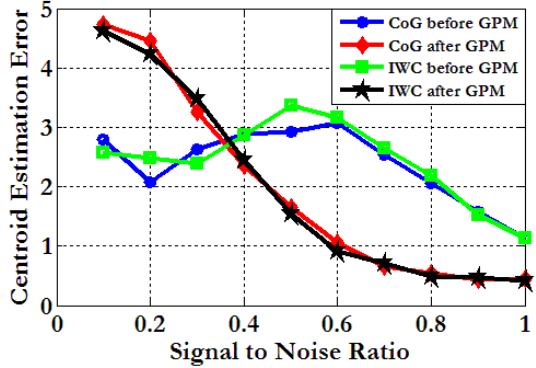

**Figure 12 Improvement in CoG and IWC methods after noise removal**

The GPM algorithm can be used in many ways. It can be used directly for centroid determination by applying CoG, IWC.

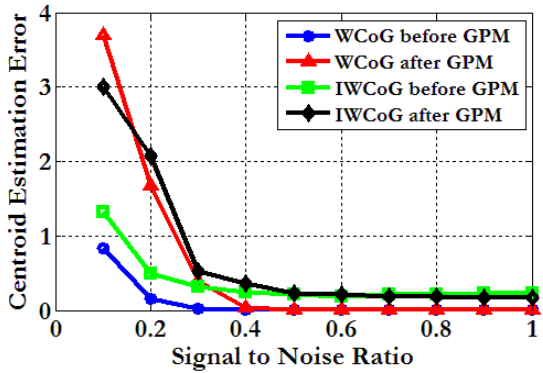

**Figure 13 Performance of GPM on WCoG and IWCoG**

WCoG and IWCoG perform better even without the application of GPM algorithm at low SNR and very small shift in the spot as shown in Fig. 13. But for large shift in the spots, the performance of WCoG is worst in the absence of GPM algorithm [10]. The performance of WCoG can be enhanced by applying GPM algorithm and use it as a centroid pre-calculator. The weighting functions for WCoG are evaluated from the pre-calculated centroid position. The improvement in the centroiding accuracy for WCoG when the spot shift is 5 pixels is shown in Fig. 14. The performance of IWCoG remains the same even for large shift in the spots. Problem arises for IWCoG only when large noise effects occur very close (closer by more than half-width of the spot) to the actual spot position. Since large noise effects occur randomly in time, it is advantageous to have GPM algorithm which can nearly point out the centroid position prior to a more precise estimation using IWCoG.

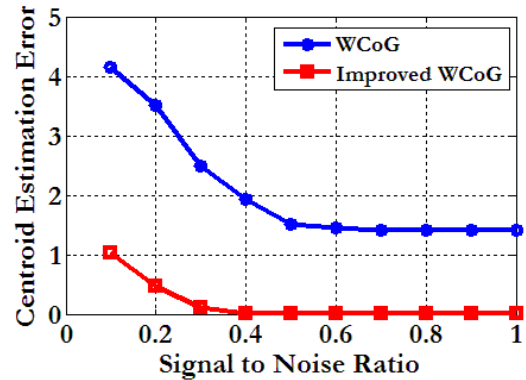

**Figure 14 Improved WCoG for 5 pixel shift in the spot**

## 6. CONCLUSIONS

Wavefront reconstruction accuracy is a critical factor in high resolution ground based astronomical imaging through turbulence. Centroiding contributes to most of the wavefront reconstruction error in Shack Hartmann sensor based adaptive optics system with readout, background noise and strong scintillations. Using an effective denoising technique based on thresholded Zernike reconstructor, the small scale noise features in the spot pattern images are removed. At very low signal to noise ratio, it was shown that taking the advantage of the shape of the spot can improve the centroiding accuracy of CoG, IWC, WCoG and IWCoG algorithms. In the case of laser guide star, an elongated spot can be used for pattern matching.